# MOONRISE: The Main MOONS GTO Extragalactic Survey

Roberto Maiolino[1,2,3]
Michele Cirasuolo[4]
José Afonso[5,6]
Franz E. Bauer[7,8,9]
Rebecca Bowler[10]
Olga Cucciati[11]
Emanuele Daddi[12]
Gabriella De Lucia[13]
Chris Evans[14]
Hector Flores[15]
Adriana Gargiulo[16]
Bianca Garilli[16]
Pascale Jablonka[15,17]
Matt Jarvis[10]
Jean-Paul Kneib[17]
Simon Lilly[18]
Tobias Looser[18]
Manuela Magliocchetti[19]
Zhongyi Man[20]
Filippo Mannucci[21]
Sophie Maurogordato[22]
Ross J. McLure[23]
Peder Norberg[24]
Pascal Oesch[25,26]
Ernesto Oliva[21]
Stéphane Paltani[25]
Ciro Pappalardo[5,6]
Yingjie Peng[20]
Laura Pentericci[27]
Lucia Pozzetti[11]
Alvio Renzini[28]
Myriam Rodrigues[10,15]
Frédéric Royer[15]
Steve Serjeant[29]
Leonardo Vanzi[7]
Vivienne Wild[30]
Gianni Zamorani[11]

[1] Kavli Institute for Cosmology, University of Cambridge, UK
[2] Cavendish Laboratory, University of Cambridge, UK
[3] Department of Physics and Astronomy, University College London, UK
[4] ESO
[5] Instituto de Astrofísica e Ciências do Espaço, Universidade de Lisboa, Portugal
[6] Departamento de Física, Faculdade de Ciências, Universidade de Lisboa, Portugal
[7] Instituto de Astrofísica, Pontificia Universidad Católica de Chile, Santiago, Chile,
[8] Millennium Institute of Astrophysics (MAS), Santiago, Chile
[9] Space Science Institute, Boulder, USA
[10] Department of Physics, University of Oxford, UK
[11] INAF – Osservatorio di Astrofisica e Scienza dello Spazio di Bologna, Italy
[12] CEA, IRFU, DAp, AIM, Université Paris-Saclay, Université Paris Diderot, France
[13] INAF – Osservatorio Astronomico di Trieste, Italy
[14] UK Astronomy Technology Centre, Edinburgh, UK
[15] GEPI, Observatoire de Paris, PSL University, CNRS, Meudon, France
[16] INAF, IASF-MI, Milano, Italy
[17] Physics Institute, Laboratoire d'Astrophysique, EPFL, Switzerland
[18] Department of Physics, ETH Zurich, Switzerland
[19] INAF – IAPS, Roma, Italy
[20] Kavli Institute for Astronomy and Astrophysics, Peking University, China
[21] INAF – Osservatorio Astrofisico di Arcetri, Firenze, Italy
[22] Université Côte d'Azur, Observatoire de la Côte d'Azur, CNRS, Nice Cedex 4, France
[23] Institute for Astronomy, University of Edinburgh, Royal Observatory, Edinburgh, UK
[24] Department of Physics, Durham University, UK
[25] Department of Astronomy, University of Geneva, Versoix, Switzerland
[26] Cosmic Dawn Center (DAWN), Copenhagen, Denmark
[27] INAF – Osservatorio Astronomico di Roma, Monte Porzio Catone, Italy
[28] INAF – Osservatorio Astronomico di Padova, Italy
[29] School of Physical Sciences, The Open University, Milton Keynes, UK
[30] School of Physics and Astronomy, University of St Andrews, UK

**The MOONS instrument possesses an exceptional combination of large multiplexing, high sensitivity, broad simultaneous spectral coverage (from optical to near-infrared bands), large patrol area and high fibre density. These properties provide the unprecedented potential of enabling, for the very first time, SDSS-like surveys around Cosmic Noon ($z$ ~ 1–2.5), when the star formation rate in the Universe peaked. The high-quality spectra delivered by MOONS will sample the same nebular and stellar diagnostics observed in extensive surveys of local galaxies, providing an accurate and consistent description of the evolution of various physical properties of galaxies, and hence a solid test of different scenarios of galaxy formation and transformation. Most importantly, by spectroscopically identifying hundreds of thousands of galaxies at high redshift, the MOONS surveys will be capable of determining the environments in which primeval galaxies lived and will reveal how such environments affected galaxy evolution. In this article, we specifically focus on the main Guaranteed Time Observation (GTO) MOONS extragalactic survey, MOONRISE, by providing an overview of its scientific goals and observing strategy.**

## An unprecedented discovery space with MOONS

The unique observing capabilities of MOONS (see for example, Figure 1 and Cirasuolo et al., p. 10) make it the optimal match to the Sloan Digital Sky Survey (SDSS) at high redshift and specifically around the Cosmic Noon, i.e., the peak of cosmic star formation, at $z$ ~ 1–2.5. Indeed, prominent optical nebular lines, typical of star-forming galaxies and active galactic nuclei (AGN), such as H$\alpha$ and [OIII] 5007 Å are detectable with MOONS out to $z$ = 1.74 and 2.6, respectively, while other bluer optical nebular diagnostics such as [OII] 3727 Å are already in the MOONS band at $z$ = 0.7 and observable up to $z$ = 3.8 (Figure 2). Key optical stellar features used to characterise the stellar populations are also in the MOONS band over this broad redshift range and will also enable the identification of passive galaxies at $z$ ~ 1–2 (Figure 2).

These are all nebular and stellar diagnostics that have been successfully used to characterise hundreds of thousands of galaxies in the local universe by extensive optical spectroscopic surveys such as SDSS and Galaxy And Mass Assembly (GAMA). Accessing these spectral features will also enable MOONS to spectroscopically measure the redshifts of galaxies in the so-called redshift desert ($z$ ~ 1.5), right at the peak of the cosmic star formation rate density, where the sparseness of spectral features has generally hampered the capability of optical spectrometers to identify galaxies. Of course, the broad spectral coverage of MOONS will also allow the investigation of large samples of very distant galaxies, around the epoch of reionisation, especially by observing Lyα and other transitions in the ultraviolet rest frame. In addition to the spectral coverage and sensitivity, the combination of high multiplexing and high density of fibres on sky will enable MOONS to identify a broad range of galaxy environments, from clusters to groups, filaments, and voids (Figure 3), and will unambiguously determine how galaxy properties depend on the environment in which they live.

Within the MOONS GTO, 190 nights are dedicated to the extragalactic survey MOONRISE (MOONS Redshift-Intensive Survey Experiment). With MOONRISE we expect to obtain key spectroscopic information for a few hundred thousand galaxies, possibly up to about half a million galaxies at $0.9 < z < 2.6$, as well as for a few thousand galaxies around the epoch of reionisation ($z \sim 6$–$8$). Such large statistics will also provide an unprecedented test of assumptions embedded in various cosmological simulations and models. Indeed, different models and simulations implement various physical processes in different ways in order to reproduce galaxy evolution and therefore they predict different galaxy properties in the early phases of their formation, which can be tested with MOONRISE's with unprecedented statistics.

Together with the MOONRISE GTO programme, during the first ten years of operation additional open time surveys will expand the legacy of MOONS to include millions of galaxy spectra spanning even broader redshift intervals and sampling a broader parameter space. In the following, we focus on the science goals and strategy of the MOONRISE survey, bearing in mind that this is only a sample of what MOONS will be able to deliver over the longer term.

## MOONRISE science goals

It is impossible to present an exhaustive list of the various cutting-edge science goals of the MOONRISE survey in the limited space available here. In the following, we outline a representative sample of the primary science aims of the survey.

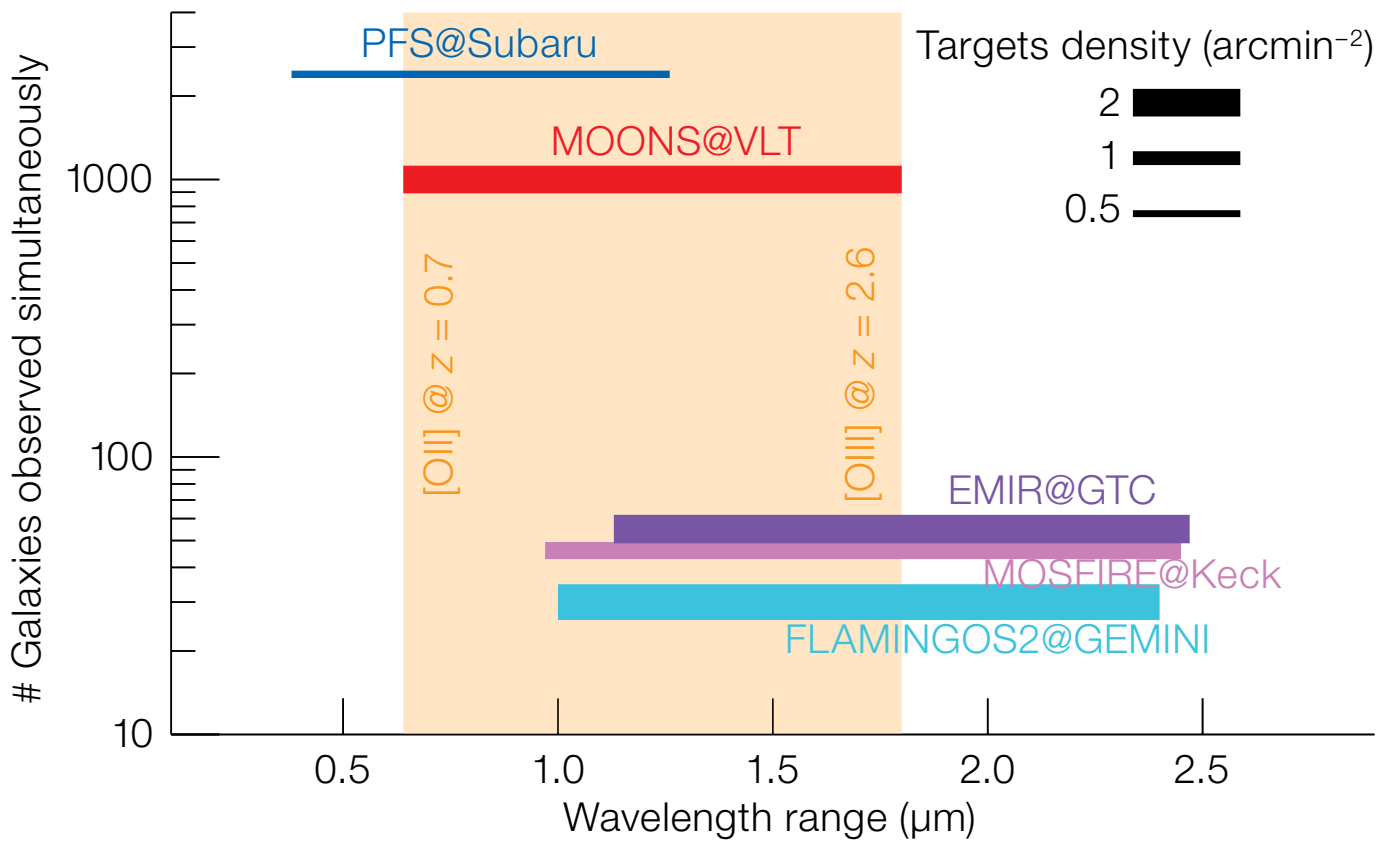

Figure 1. Comparison between MOONS and other near-infrared multi-object spectrographs at ground-based 8–10-m telescopes in terms of wavelength range, multiplexing and maximum target density that can be observed in a single pointing.

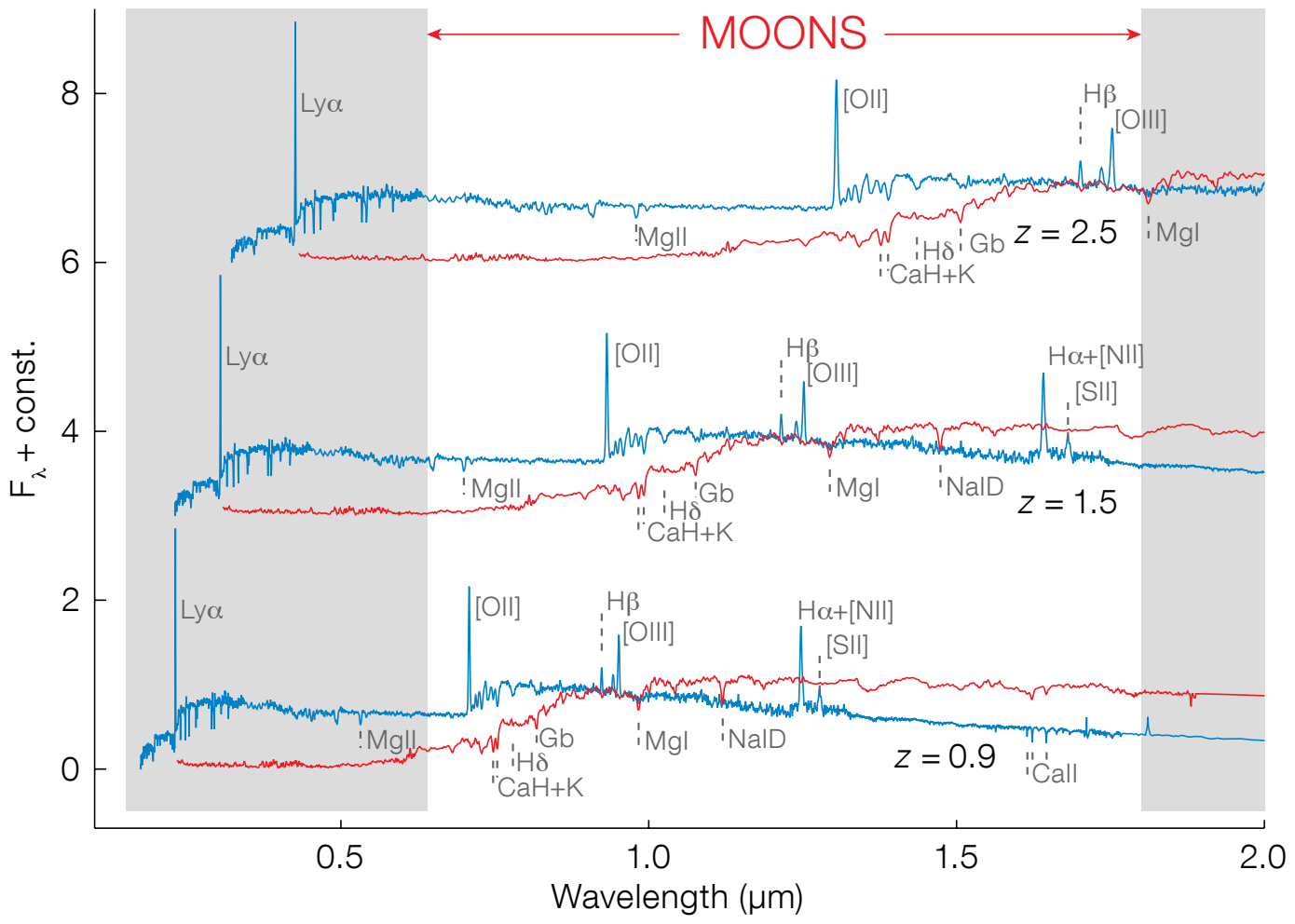

Figure 2. Examples of star-forming (blue) and passive (red) galaxy spectra shifted to three representative redshifts that will be targeted by the MOONRISE survey (Table 1), illustrating the observability of some of the primary nebular and stellar rest-frame optical features.

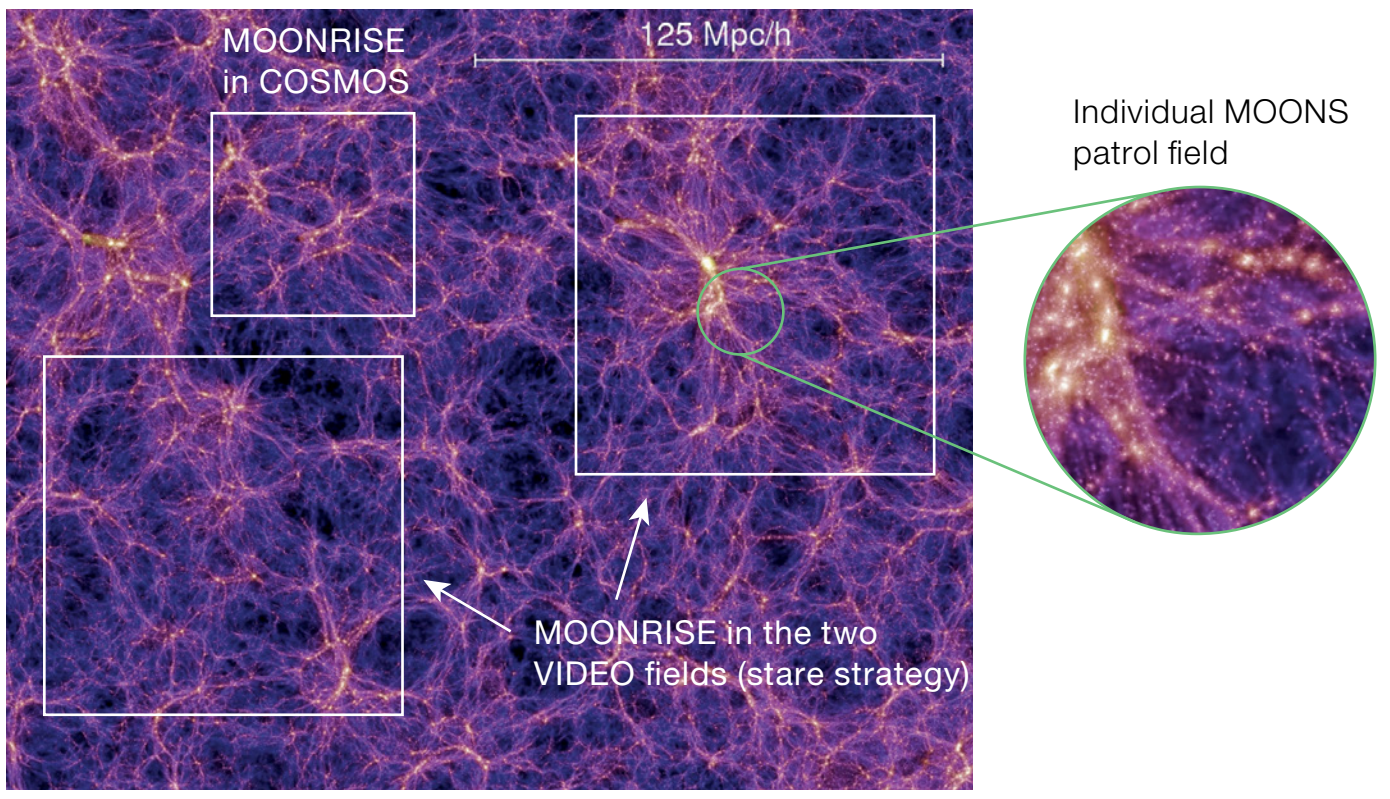

Figure 3. Example of different environments that will be sampled by the MOONRISE survey at a representative redshift slice around $z = 1.4$ by exploiting the large-scale dark matter distribution obtained from the Millennium simulation (Springel et al., 2005).

– The metallicity evolution of galaxies. By measuring multiple nebular transitions in galaxies, spanning more than three orders of magnitude in mass ($10^{8.5} < M_*/M_\odot < 10^{11.7}$) and more than three orders of magnitude in the star formation rate (SFR), it will be possible to measure the gaseous metallicity for hundreds of thousands of galaxies, enabling us to solidly assess the evolution of the metallicity scaling relations. In particular, the redshift evolution (or lack thereof) of the mass-metallicity relation (for example, Troncoso et al., 2014; Kashino et al., 2019) and of the Fundamental Metallicity Relation (the relation between mass, SFR and metallicity; Mannucci et al., 2010) are still open and hotly debated problems. These are also key correlations predicted by various cosmological simulations (Maiolino & Mannucci, 2019). MOONS will measure these scaling relations and their scatter at $z \sim 1$–2 with unprecedented accuracy (Figure 4). Stellar metallicities will also be measured for tens of thousands of galaxies individually and for hundreds of thousands of galaxies through stacking, which will provide further tight constraints to galaxy evolutionary scenarios (for example, Trussler et al., 2019; Cullen et al., 2019).

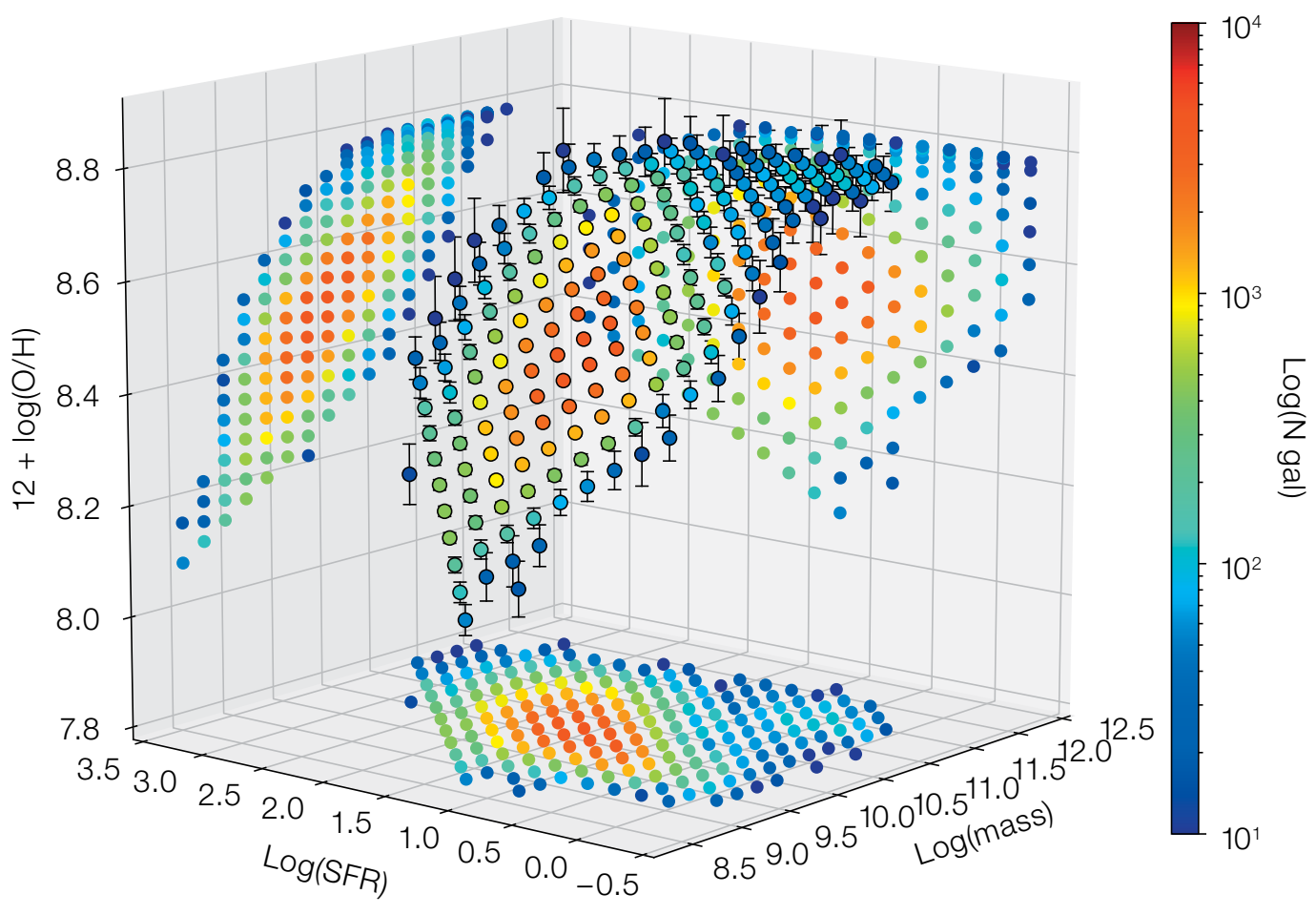

**Figure 4.** Three-dimensional diagram illustrating the capability of the MOONRISE survey to determine the shape of the fundamental metallicity relation — the relation between metallicity, mass and star formation rate — at $z \sim 1.5$, by measuring the metallicity of hundreds of thousands of galaxies. The colour coding gives the number of galaxies in each bin of mass and SFR. Galaxies are only included in this plot when metallicity can be determined by at least two independent indicators.

– Chemical abundances. Nebular and stellar features will enable us to investigate the relative abundance of chemical elements produced on different timescales (for example, N/O, $\alpha$/Fe) for several thousands of galaxies and, through stacking, for up to hundreds of thousands of galaxies at high redshift. These relative chemical abundances will provide key information on the evolutionary timescales of primeval galaxies, on the efficiency of star formation, and on gas flows in and out of galaxies (Vincenzo et al., 2016; Thomas et al., 2010; De Lucia et al., 2017; Pipino et al., 2008).

– Active galactic nuclei and black hole demographics. MOONS's broad spectral coverage will allow us to trace the excitation diagnostics (for example, the BPT diagnostic [see Baldwin, Phillips & Terlevich, 1981], and high-ionisation species, such as HeII) across a broad redshift range encompassing the Cosmic Noon. It is expected that the MOONRISE survey will identify at least a few tens of thousands of type 2 AGN, including those which are completely absorbed in X-rays (Compton thick), down to Seyfert-like luminosities. This will enable us to unambiguously assess if and how black hole accretion correlates with star formation, with galaxy interactions or with the galaxy transition to quiescence. Comparing these results with cosmological simulations will allow us to understand the reciprocal role played by black hole accretion in galaxy formation, and vice versa. MOONRISE will also identify a few thousand type 1 AGN for which we will be able to measure the black hole mass by using the same methodology as in local type 1 AGN (virial relations calibrated on the broad component of H$\beta$ and/or H$\alpha$). As a consequence, it will be possible to obtain an unprecedented census of black holes at high redshift, down to low masses (for those Seyfert 1's accreting close to their Eddington limit).

– Kinematics. The excellent spectral resolution of MOONS ($R \sim 7000$ in the $H$ band) will enable a detailed analysis of the velocity profile of the stellar and nebular lines that is much more accurate than what can be achieved by the more modest resolution of SDSS or GAMA. The velocity dispersion will provide basic information on the dynamical mass down to low-mass systems and it will make it possible to explore the cosmic evolution of dynamical scaling relations such as the fundamental plane. The profile of the nebular emission and absorption lines (for example, [OIII], H$\alpha$, NaDI) will allow us to trace the demographics of galactic outflows, their physical properties and their relation with star formation rate, with galaxy mass, and with the presence of an AGN, which are key phenomena responsible for suppressing or possibly boosting star formation in galaxies (for example, Cicone et al., 2016; Concas et al., 2017, 2019; Nedelchev et al., 2019; Gallagher et al., 2019).

– Passive galaxies, galaxy transformation and star formation histories. Thanks to the high spectral resolution of MOONS it will be possible to mask the strong OH sky lines and rebin the spectrum to a resolution optimised to detect the continuum of each galaxy while preserving the information on the various stellar features (Figure 5). The spectroscopic characterisation of the stellar continuum along with the properties of the nebular emission lines (either absent or not associated with star formation) will unambiguously confirm candidate passive galaxies at high redshift (Citro et al., 2016; Merlin et al., 2019). The cosmic evolution of the fraction of passive galaxies is one of the most fundamental aspects required for the understanding of galaxy evolution, as it constrains the onset and mechanisms of star formation quenching in galaxies. Detecting the stellar continuum with high S/N will enable us to constrain the age and star formation history of galaxies (for example, Trussler et al., 2019; Carnall et al., 2019; Wild et al., 2020). This will be particularly interesting for galaxies in the so-called green valley — galaxies in transition from the star formation Main Sequence to quiescence — as it will enable us to trace the timeframe of the quenching process in galaxies. With the observing strategy envisaged for MOONRISE we expect to detect the continuum of thousands of individual passive galaxies down to $H_{AB} < 22$ and, through stacking, of tens of thousands of passive galaxies down to at least $H_{AB} < 23$. In the case of star-forming and green-valley galaxies (for which the detection of — some — nebular emission lines will facilitate the determination of the spectroscopic redshift) stacking will be possible down to $H_{AB}$ = 25.

– The role of the environment. Huge statistics, high target density, near-infrared spectral coverage and high sensitivity make MOONS the optimal machine to determine and characterise the environment in which high-$z$ galaxies live and to tackle one of the key open problems in galaxy evolution, namely if, how and when environment has affected galaxy evolution and galaxy properties.

The MOONRISE survey is expected to explore the environment of galaxies at $z$ ~ 1–2 over four orders of magnitude in terms of galaxy overdensity (from $\log(1+\delta) \sim -2$ to $\log(1+\delta) > 2$), i.e., from voids to (proto-)clusters. The observing strategy will allow us to identify groups, derive their halo masses and disentangle central galaxies from their satellites. It will also be possible to trace filaments connecting tens of (proto-)clusters. By combining this environmental outcome with the information on galactic properties obtained with the same MOONS spectra it will be possible to assess with unprecedented accuracy whether different environments accelerate or inhibit galaxy evolution and galaxy transformation, and to determine the different evolutionary paths of central and satellite galaxies. As an example, Figure 6 illustrates how the MOONRISE survey will be able to discriminate between the effects of environmental and mass quenching at $z$ ~ 1.5, comparable to local SDSS studies (Peng et al., 2010). Moreover MOONRISE will give access to the phase-space distribution from the core to the outskirts of clusters/groups, which will provide an unprecedented view of the evolution of galaxy physical properties with respect to their infall history in their host, for a variety of halo masses.

– Galaxies at the dawn of the Universe. While performing the primary survey of galaxies at Cosmic Noon, a fraction of the fibres will be allocated to observing in depth galaxies probing the Cosmic Dawn. Specifically, we expect to obtain deep spectra of a few thousand galaxies at $z$ ~ 5–10. The MOONS sensitivity, combined with its spectral resolution, is expected to provide key information on the visibility and shape of the Ly$\alpha$ line, which in turn provides precious information on the amount of neutral gas in the intergalactic medium and on the escape fraction of ionising photons. MOONS will also have enough spectral coverage to measure various other lines in the ultraviolet rest frame portion of the spectra (CIV, HeII, CIII], NV), which will provide key information to identify primeval AGN and to determine metal enrichment, hardness of the ionising radiation and ionisation parameter in these primeval systems; these are all key quantities that will be compared with predictions of galaxy formation scenarios and reionisation models.

## Survey strategy

The GTO team has been working on a reference survey that is expected to deliver the statistics, depth and area appropriate to achieving the goals discussed above, as well as tackling many more science cases.

We will target three main fields: the Cosmic Evolution Survey field (COSMOS) and the two fields XMM-LSS and ECDFS from the VISTA Deep Extragalactic Observations (VIDEO) Survey (see Jarvis et al., 2013). Together with the Galactic survey, these fields will enable an efficient scheduling of the GTO. We plan to observe 1 square degree in COSMOS (part of the area covered more extensively in multiple bands, see Laigle et al., 2016 and Nayyeri et al., 2017), and 3 to 6 square degrees in the other two fields depending on the observing mode, as discussed in the following.

Targets will be selected from multi-band optical to near-infrared photometry in the three main (photometric) redshift ranges listed in Table 1 in which most of the primary optical nebular lines, as well as primary optical stellar features, are observable in the MOONS band (Figure 2) and also avoiding significant overlap with other MOS spectrographs at 8-m telescopes (for example, PFS at Subaru can observe Hα out to $z = 0.9$). As some of the science cases require high completeness over a mass-limited sample while other science goals require large statistics on a magnitude-limited sample, we have applied a double selection whereby in each redshift bin galaxies are selected through their mass *or* their magnitude as listed in Table 1, corresponding to about 6000 targets per MOONS patrol field, allowing an efficient allocation of the 1000 fibres.

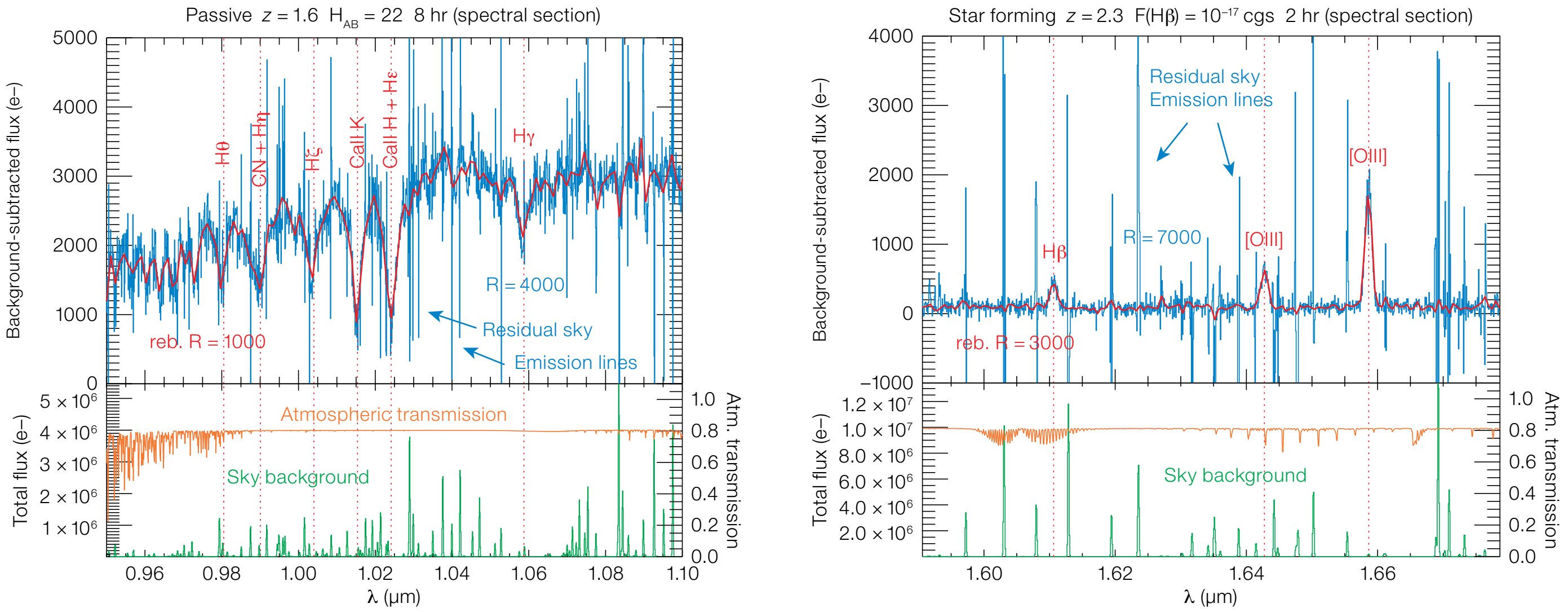

Figure 5. Portions of MOONS simulated spectra of a passive galaxy ($H_{AB}$ = 22) at $z$ = 1.6 (left) observed for 8 hours on source and a star forming galaxy ($F$(H$\beta$) = $10^{17}$ erg s$^{-1}$ cm$^{-2}$, $H_{AB}$ = 23.5) at $z$ = 2.3 (right) observed for two hours on source. The bottom panels show the total observed spectrum (green), including sky emission, and the atmospheric transmission (brown). The top panels show the background-subtracted spectrum (blue) and the same spectrum rebinned to lower resolution after masking the OH sky lines (red).

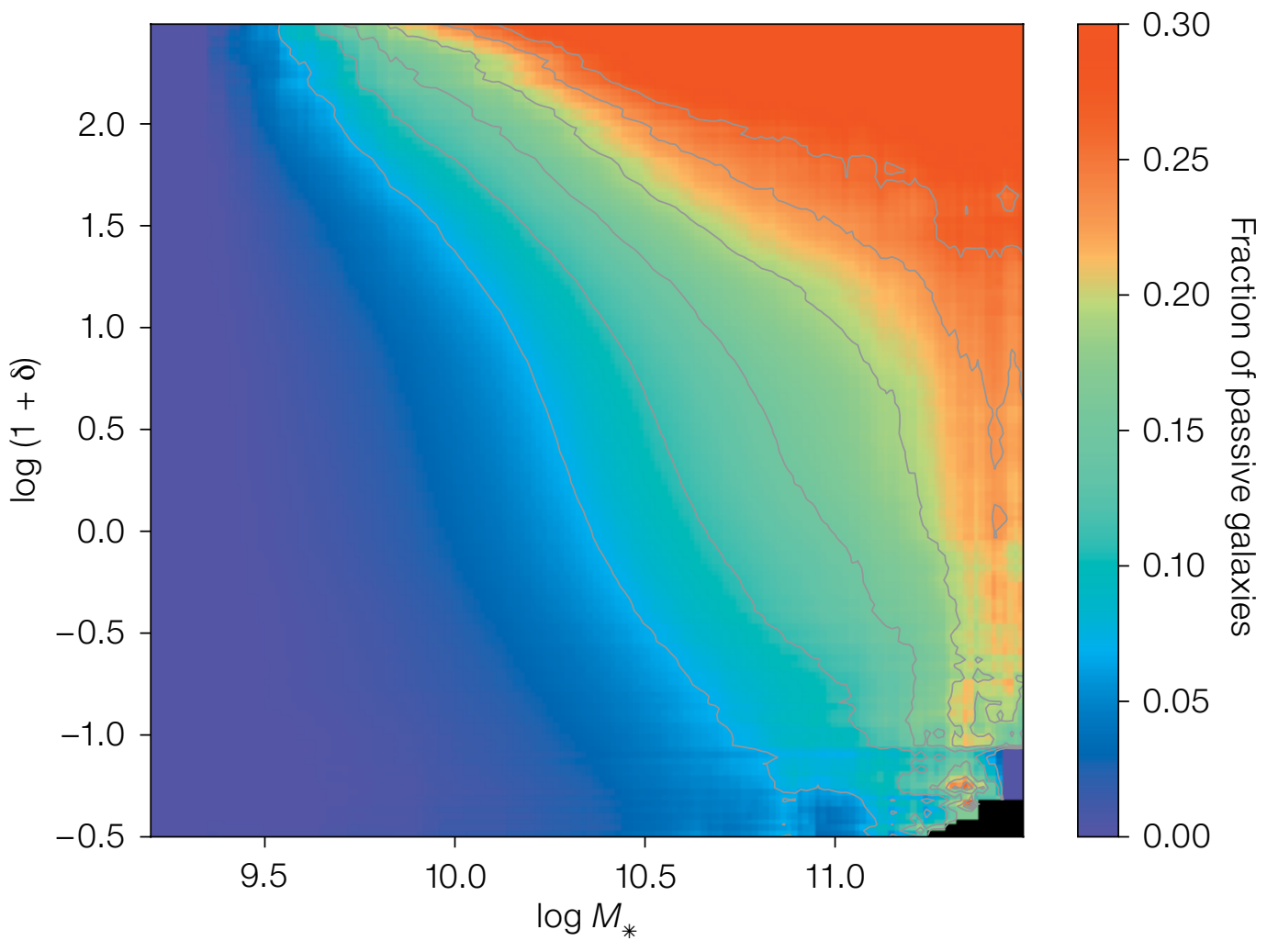

**Figure 6.** Fraction of passive galaxies as a function of galaxy overdensity and as a function of galaxy mass for satellite galaxies at $1.3 < z < 1.8$ from the simulated MOONRISE survey at $H_{AB} < 23$ (assuming completeness of 70% over an area of 6 square degrees).

High completeness is required for many science cases discussed above, especially for properly assessing the environment, and can be achieved by observing the same pointing several times with different fibre configurations. Obviously, with this strategy the efficiency of allocating fibres to new targets gradually decreases (as the density of unobserved targets rapidly decreases), therefore a compromise between survey efficiency and completeness has to be achieved. In the current survey design we aim at achieving a completeness of 80% in the COSMOS field and 70% in the two VIDEO fields. The number of repeated observations required to achieve these levels of completeness will depend on whether observations have to be performed in Xswitch mode (i.e., ~ 400 pairs of fibres allocated to targets and sky for nodding observations) or in Stare mode (i.e., ~ 900 fibres allocated to targets while the background is sampled through a few tens of fibres scattered over the patrol field; see Cirasuolo et al., p. 10). These allocation efficiencies are conservative and take into account potential limits associated with the fibre allocation algorithms. Whether the Stare mode gives an acceptable background subtraction or we need to resort to using the more traditional but less efficient Xswitch nodding mode will only be assessed at the telescope.

The detection of nebular emission lines in star-forming galaxies, even in systems as faint as $H_{AB} \sim 25$ (in the mass-limited samples), can be achieved much more quickly than for the stellar continuum in passive galaxies. Our baseline strategy is to allocate 25% of the fibres to passive galaxies for each pointing, which will be repeatedly observed for a total of 8 hours on source, while at the same time all other fibres are reallocated to different star-forming galaxies, which will be observed with exposures of 1 or 2 hours (the relative fraction of fibres allocated to passive and star-forming galaxies stems from the relative density on sky of these two populations). With this strategy, in COSMOS we will observe (in Xswitch mode) ~ 480 passive galaxies and ~ 4350 star-forming galaxies in each MOONS pointing (resulting in the desired 80% completeness) for a total observing time of 38 nights, including overheads and allowance for bad weather. These figures nearly double if the Stare observing mode can be adopted. In the VIDEO fields we will adopt fewer repeated observations resulting in a lower completeness of 70%. Given the multiple passes in each pointing to reach completeness, a subset of targets will be observed multiple times, with integrations up to 40 hours on source.

As mentioned earlier, a few tens of fibres in each pointing will be allocated to candidate galaxies at $z > 5$, identified via broad-band photometry through the Lyman-break technique or as Lyman-α emitter candidates identified in narrow-band surveys. Each of these high-*z* galaxies will be observed with integrations of 8 hours each. A similar allocation of a few tens of fibres per field is also being considered for X-ray-selected AGN, while additional AGN will be identified among the magnitude/mass selected samples through their nebular line diagnostics.

| Redshift range | Main spectral features | Selection | Number of galaxies* Xswitch (4 square degrees) | Stare (7 square degrees) |
|---|---|---|---|---|
| $0.9 < z < 1.1$ | [OII], Hβ, [OIII], Hα, [NII], [SII] | $H_{AB} < 23$ or | 33 900 | 75 300 |
| | CaH+K, Hδ, Gb, Mgb, NaID, CaII | $\log(M_*) > 9.5$ | 12 900 | 28 500 |
| $1.2 < z < 1.7$ | [OII], Hβ, [OIII], Hα, [NII], [SII] | $H_{AB} < 23.5$ or | 88 700 | 197 100 |
| | MgII, CaH+K, Hδ, Gb, Mgb, NaID | $\log(M_*) > 9.5$ | 13 700 | 30 500 |
| $2.0 < z < 2.6$ | [OII], Hβ, [OIII] | $H_{AB} < 24$ or | 54 500 | 121 100 |
| | MgII, CaH+K, Hδ, Gb, Mgb | $\log(M_*) > 10$ | 2100 | 4700 |
| $5 < z$ | Lyα, NV, HeII, CIV, CIII] | $H_{AB} < 26$ | 2000 | 4500 |
| Total | | | 207 800 | 461 700 |

**Table 1.** Summary of the MOONRISE survey design. The blue numbers represent star-forming galaxies and AGN and the red numbers are passive galaxies.

The total number of galaxies expected to be observed in MOONRISE is given in Table 1.

We would, however, like to emphasise that the survey design is still being optimised by taking into account the analysis of photometric samples and mock samples that are being refined, as well as the optimisation of the fibre allocation software and the eventual performance of the instrument on sky.

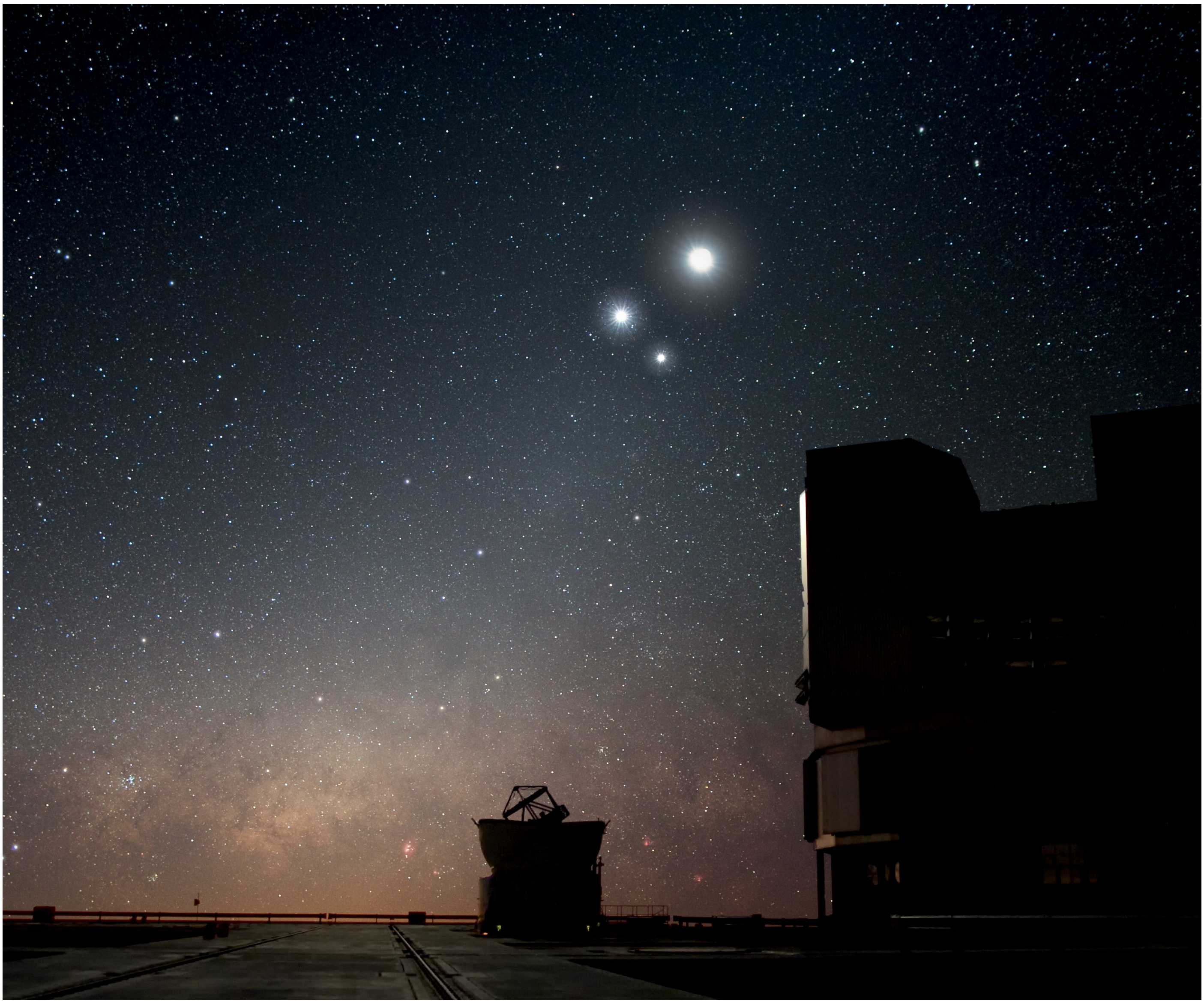

ESO/Y. Beletsky

A celestial conjunction, involving the Moon, Venus and Jupiter, shines above one of the VLT Unit Telescopes and an Auxiliary Telescope.